\documentclass[aps,prc,superscriptaddress,twocolumn,showpacs,nofootinbib]{revtex4-1}
\usepackage[utf8]{inputenc}
\usepackage[T1]{fontenc}
\usepackage{graphicx}   
\usepackage{latexsym}   
\usepackage{enumerate}
\usepackage{rotating,booktabs,multirow}
\usepackage{amsmath}
\usepackage{amsfonts}
\usepackage{amssymb}
\usepackage{multirow}
\usepackage{ulem}
\usepackage{bm}
\usepackage{subfigure}
\usepackage{xcolor}
\usepackage[colorlinks=true,linktocpage=true,linkcolor=blue,citecolor=blue,allcolors=blue]{hyperref}

\begin{document}
\title{Unshadowing the constituent quark number scaling of harmonic flow in heavy-ion collisions}

\author{Tom Reichert}
\affiliation{Theoretical Physics Department, CERN, 1211 Geneva 23, Switzerland}
\affiliation{Department of Physics, Duke University, Durham, NC 27708, USA}

\author{Iurii Karpenko}
\affiliation{Faculty of Nuclear Sciences and Physical Engineering, Czech Technical University in Prague, B\v{r}ehov\'a 7, 11519 Prague 1, Czech Republic}

\begin{abstract}
Constituent quark number scaling of elliptic flow has been proposed as one key observable to identify the phase transition or the absence of the Quark-Gluon Plasma (QGP) in heavy-ion collisions. At the fixed target program at RHIC the STAR collaboration has recently reported that NCQ scaling breaks when decreasing the collision energy from $\sqrt{s_\mathrm{NN}} = 4.5$ to $3.0$ GeV. However, the generation of elliptic flow is dominated by a highly intricate interplay of spectator shadowing, squeeze-out and geometry dependent hadron emission governed by their cross sections. Therefore in this article we will disentangle the shadowing contribution from the harmonic flow signal of the particle emitting source, effectively ``unshadowing'' the source. We introduce Fourier coefficients that quantify the azimuthal absorption rate of hadrons decoupling from the system. We benchmark the derived results using a toy model based on a ballistic Glauber description of the penetrating nuclei and calculate how shadowing qualitatively alters the constituent quark number scaling of the hadron emitting source.
The results are thus relevant for interpreting recent STAR measurements as well as the upcoming measurements by CBM at FAIR.
\end{abstract}


\maketitle

\section{Introduction}
Strongly interacting matter, in theory described by Quantum Chromo Dynamics (QCD), is nowadays studied in the worlds largest particle accelerators. These span a huge range of collision energies stretching from the GeV regime at SIS at GSI, over RHIC at BNL to the TeV regime at the LHC at CERN. At the highest collision energies, or equivalently at high temperatures and vanishing baryon chemical potential, nuclear matter undergoes a crossover transition to a quasi-free state of quarks and gluons, known as the Quark Gluon Plasma (QGP) \cite{Cabibbo:1975ig}. 

While at vanishing chemical potential lattice QCD (lQCD) calculations show a crossover transition \cite{Borsanyi:2010bp,Bazavov:2011nk,Bazavov:2017dus}, ab initio calculations at finite baryon chemical potential become unfeasible. It is, however, hypothesized that with decreasing temperature and increasing baryon chemical potential the phase transition becomes a first order transition ending in a critical end point. Such a point is characterized by diverging correlation length and critical fluctuations and is of key interest of current and upcoming experiments, e.g. STAR-FXT at RHIC and CBM at FAIR \cite{Sorensen:2023zkk,Agarwal:2025ezo,Messchendorp:2025men}. Recent theoretical efforts mostly converge on a region of $T^c \approx 80-120$ MeV and $\mu_B^c \approx 450 - 650$ MeV \cite{Fischer:2014ata,Fu:2019hdw,Gao:2020fbl,Gunkel:2021oya,Hippert:2023bel,Basar:2023nkp,Clarke:2024seq,Sorensen:2024mry,Shah:2024img,Ecker:2025vnb}, with a few exceptions favoring larger chemical potentials \cite{Eser:2023oii,Steinheimer:2025hsr}. 

The existence of the QGP at the highest collision energies is evident from several observables such as strangeness enhancement, jet quenching, or the near perfect liquidity. However, at lower collision energies it is not straightforward to conclude from experimental data whether a QGP has been created during the collision or not. One promising observable is constituent quark number scaling (NCQ scaling). Under the assumption of a quark coalescence picture for hadron formation, the elliptic flow of a hadron $v_2^h$ scales with the elliptic flow of its constituent quarks $v_2^q$ at $N_q$ times the momentum, i.e. $v_2^h(p_\mathrm{T}^h) = N_q v_2^q(p_\mathrm{T}^q)$ with $p_\mathrm{T}^h = N_q p_\mathrm{T}^q$. Such a scaling has been observed by the STAR collaboration across various collision energies \cite{STAR:2012och,STAR:2013cow,STAR:2013ayu,STAR:2015rxv} and it is also known from cluster production, which scale with their respective constituents (protons and neutrons), from the HADES experiment \cite{HADES:2020lob,HADES:2022osk}, although ALICE has reported a breaking of constituent quark number scaling \cite{ALICE:2014wao}. The observation of such a kind of scaling has been widely accepted as evidence for partonic collectivity, although hadronic transport models also predict similar scaling relations \cite{Lu:2006qn} without partonic phase. Violations of NCQ scaling at intermediate BES energies have so far been attributed to baryon stopping and a different partonic flow profile of produced and transported quarks \cite{Dunlop:2011cf}, however, this explanation cannot capture spectator shadowing present at low collision energies.

With decreasing collision energy, the bypassing baryon current (i.e. the spectators) does not decouple instantly as at the highest energies. Instead, the spectator is present during the course of the collision, effectively blocking the free expansion of the fireball in the reaction plane through absorption of transverse momentum. It thus directly alters the measureable azimuthal particle distribution, shadowing the angular distribution of the source. Such an effect, known as spectator shadowing, leads e.g. to negative elliptic flow of protons \cite{Reichert:2024ayg}. It is a priori not clear how constituent quark number scaling would show up in experimental data at low-GeV heavy-ion collisions under the presence of spectator shadowing.

In their fixed target run, the STAR collaboration has recently pushed the investigation of constituent quark number scaling down to such small collision energies and large net baryon density and published measurements of the elliptic flow of several hadron species in the STAR-FXT energy regime \cite{STAR:2021yiu,STAR:2025owm}. Therein they claim to have observed a breaking of (NCQ) scaling of the hadron $v_2$ below $\sqrt{s_\mathrm{NN}} \approx 4.5$ GeV. This breaking has been interpreted as ``disappearance of partonic collectivity'' \cite{STAR:2021yiu}. 

In this article we are therefore going to derive a method to systematically subtract the shadowing contribution from the measured elliptic flow signal. This hence allows to access the azimuthal particle distribution of the emission source and investigate whether the source shows features of constituent quark number scaling. The method is based on a Fourier decomposition of the transverse shadowing strength given by the absorption probability or optical depth. The article is of interest for experiments studying few-GeV heavy-ion collisions (e.g. HADES, STAR and the upcoming CBM), but the presented method is valid independent of the collision energy.



\section{Unshadowing harmonic flow}

\subsection{NCQ scaling from quark coalescence}
The formation of hadrons from a partonic phase is generally a non-trivial and not yet fully understood process. However, much progress on hadronization has been made in effective models of QCD and in certain limits or momentum ranges. While hadron formation at intermediate momenta has a strong contribution from quark fragmentation, hadron formation at low transverse momenta (in transport models) mainly happens through quark coalescence\footnote{In fluid-dynamic models, the partonic phase is implicitly encoded via the equation of state, and hadronization is typically assumed to happen in the fluid phase/stage. Produced hadrons are sampled according to the so-called Cooper-Frye prescription, which assumes (democratic) emission from thermalized and chemically equilibrated medium. As such, coalescence mechanism for hadron production is replaced by the assumption of complete thermal and chemical equilibration at the moment when the dense medium decouples, and it effectively leads to very similar observed hadron distributions.}. Quark coalescence models generally assume that the phase space (or Wigner) distribution of a hadron is given by the product of the phase space (or Wigner) distributions of its constituents times a coalescence probability or volume. Assuming that all (anti-)quarks have a common distribution function, quark coalescence leads to the following hadron distribution
\begin{align} \label{eq:coalescence}
    \frac{\mathrm{d}N^h}{\mathrm{d}x_h^3 \mathrm{d}p_h^3} &\propto \int \rho(\Delta x, \Delta p) \prod\limits_{q \in h} \mathrm{d}^3x_q \mathrm{d}^3p_q \left( \frac{\mathrm{d}N^q}{\mathrm{d}x_q^3 \mathrm{d}p_q^3} \right),
\end{align}
where $\rho$ is the coalescence volume and the product runs over the constituent quarks $q$ making up hadron $h$. Eq. \eqref{eq:coalescence} then yields the following proportionality $\mathrm{d}N^h(p_\mathrm{T}) \propto \left(\mathrm{d}N^q(p_\mathrm{T}/N_q)\right)^{N_q}$, where the hadron's transverse momentum is a multiple of the partons' transverse momenta. 

Several studies have used this as a starting point to derive formulas relating the angular distribution of the partons to the angular distribution of the hadrons \cite{Nonaka:2003hx,Molnar:2003ff,Kolb:2004gi}. The most formal derivation can be found in detail in \cite{Kolb:2004gi}, however, we shortly recap the article's main idea. The azimuthal angular distribution of the valence quarks can be expanded as a Fourier series, and so can the angular distributions of the composite hadrons, i.e.
\begin{align}
    f^q(\phi) &\propto 1 + 2 \sum\limits_{n=1}^\infty v_n^q \cos(n \phi) \\
    F^h(\phi) &\propto 1 + 2 \sum\limits_{n=1}^\infty v_n^h \cos(n \phi),
\end{align}
where $f^q$ is the azimuthal distribution of partons, $F^h$ is the azimuthal distribution of hadrons and the $v_n^{q,h}$ are the $n^\mathrm{th}$ order flow harmonics or Fourier coefficients of the partons or hadrons, respectively. Assuming a coalescence picture the azimuthal distributions of mesons and baryons can then be written as squares or third powers of the quark distribution, i.e. $F_M(\phi) \propto f_q^2(\phi)$ and $F_B(\phi) \propto f_q^3(\phi)$. Now one can insert the Fourier expansion of the quark distribution, apply trigonometric theorems and rearrange all terms with common modularity $n\phi$. The flow coefficients of the hadron can then be expressed in terms of the flow coefficients of the partons. In leading order this reduces to the known and commonly applied scaling relations
\begin{align}
    v_2^M(p_\mathrm{T}) &= 2 v_2^q(p_\mathrm{T} / 2) \label{eq:leading_order_NCQ_M} \\
    v_2^B(p_\mathrm{T}) &= 3 v_2^q(p_\mathrm{T} / 3), \label{eq:leading_order_NCQ_B}
\end{align}
where higher order terms have been neglected. For the full expression, see e.g. \cite{Kolb:2004gi}.

We, again, point out that this method assumes that the partonic azimuthal angle distributions of different quark flavors are equal. This can obviously fairly easily be generalized to different quark flavors or a separation into produced and leading quarks. One should also note that the leading order expression shown in Eqs. \eqref{eq:leading_order_NCQ_M}, \eqref{eq:leading_order_NCQ_B} might not lead to exact scaling when the higher order terms become more relevant. This is e.g. seen in the scaling of the flow of light nuclei formed through coalescence from protons and neutrons, see \cite{HADES:2022osk}.

\subsection{Spectator shadowing}
The derivation has so far implicitly assumed that the transverse expansion of the formed hadrons is unblocked (or at least blocked in each direction equally strongly), i.e.\ the azimuthal angle distribution of the hadrons doesn't change after their formation. However, this assumption is not valid anymore at small center of mass energies. Here the spectators do not decouple instantly from the collision zone and instead block the free expansion of the system during their passing time. The passing time is inversely proportional to the center of mass energy $t_\mathrm{pass} \sim (\sqrt{s_\mathrm{NN}})^{-1}$, which is why it vanishes at large collision energies. The shadowing strength is defined by the transverse optical depth or the hadron's probability of escape $P_\mathrm{esc}(\phi)$ along its trajectory. For a hadron formed at $(t,\mathbf{x})$ and propagating with $\mathbf{p}$ its probability of escape is given by \cite{Knoll:2008sc}
\begin{align} \label{eq:Pesc}
    P_\mathrm{esc}(\phi) &\propto \exp \left( - \int\limits_{t}^\infty \mathrm{d}t^\prime \underbrace{\sigma(\sqrt{s}) |v_\mathrm{rel}| \rho(t^\prime, x^\prime)}_{R(t^\prime, x^\prime)} \right) ,
\end{align}
where $R$ is the scattering rate, $\sigma$ the cross section, $v_\mathrm{rel}$ the relative velocity, $\rho$ is the density and $(t^\prime,x^\prime)$ are the coordinates along the particle's trajectory. This clearly makes the escape probability dependent on the azimuthal angle $\phi$ as well as the passing time allowing one to write $P_\mathrm{esc}$ formally as a Fourier series as well.
\begin{align}
    P_\mathrm{esc}(\phi) &\propto 1 + 2 \sum\limits_{n=1}^\infty p_n \cos(n \phi)
\end{align}
The shadowing coefficients (Fourier coefficients of $P_\mathrm{esc}(\phi)$), denoted by $p_n$, are therefore species dependent via the cross section. Also the azimuthal shadowing distribution will have a very strong (negative) elliptic component $p_2$ at midrapidity and a strong negative (positive) directed component $p_1$ at forward (backward) rapidity, which both are dictated by the geometry of the collision system. This already makes it evident that a comparison of hadrons with drastically different absorption cross sections on nuclear matter will be very valuable. Prime candidates are the pion, having an average mean free path of $\lambda_\mathrm{mfp} \approx 1$ fm in nuclear matter, and the $\phi$ meson, which decouples from the system nearly undisturbed.

The probability that two (three) partons form a meson (baryon) at double (three times) their momentum, which is finally detected under the azimuthal angle $\phi$ without absorption or rescattering in the bypassing spectator, i.e. the measureable harmonic flow coefficients, is thus given by the product $\mathcal{F}_M(\phi) \propto F_M(\phi) P_\mathrm{esc}(\phi) = f_q^2(\phi) P_\mathrm{esc}(\phi)$ for mesons and $\mathcal{F}_B(\phi) \propto F_B(\phi) P_\mathrm{esc}(\phi) = f_q^3(\phi) P_\mathrm{esc}(\phi)$ for baryons. From here on one can expand the Fourier series, rearrange the terms and collect common modularities (see appendix for detailed derivation). The harmonic flow coefficients of hadrons $\mathcal{V}^h_n$ are then given by $\mathcal{V}^h_n = C^h_n / (2 C^h_0)$,
where for the mesons the $C^M_{0,n}$ are
\begin{align}
    C^M_0 &= 1 + 2 \sum\limits_{i=1}^\infty \sum\limits_{j=1}^\infty v_i (v_j + 2p_j) \delta_{|-i+j|,0} \\
    &+ 4 \sum\limits_{i=1}^\infty \sum\limits_{j=1}^\infty \sum\limits_{k=1}^\infty v_i v_j p_k \delta_{|-i+j+k|,0} \nonumber \\
    &+ 2 \sum\limits_{i=1}^\infty \sum\limits_{j=1}^\infty \sum\limits_{k=1}^\infty v_i v_j p_k \delta_{|-i-j+k|,0} \nonumber 
\end{align}
and
\begin{align}
    C^M_n &= 4 v_n + 2 p_n + 2 \sum\limits_{i=1}^\infty \sum\limits_{j=1}^\infty v_i (v_j + 2p_j) \delta_{i+j,n} \\
    &+ 2 \sum\limits_{i=1}^\infty \sum\limits_{j=1}^\infty v_i (v_j + 2p_j) \delta_{|-i+j|,n} \nonumber \\
    &+ 4 \sum\limits_{i=1}^\infty \sum\limits_{j=1}^\infty \sum\limits_{k=1}^\infty v_i v_j p_k \delta_{|-i+j+k|,n} \nonumber \\
    &+ 2 \sum\limits_{i=1}^\infty \sum\limits_{j=1}^\infty \sum\limits_{k=1}^\infty v_i v_j p_k \delta_{|-i-j+k|,n} \nonumber \\
    &+ 2 \sum\limits_{i=1}^\infty \sum\limits_{j=1}^\infty \sum\limits_{k=1}^\infty v_i v_j p_k \delta_{i+j+k,n} \nonumber 
\end{align}
and for baryons the $C^B_{0,n}$ are
\begin{align}
    C^B_0 &= 1 + 6 \sum\limits_{i=1}^\infty \sum\limits_{j=1}^\infty v_i (v_j + p_j) \delta_{|i-j|,0} \\
    &+ 2 \sum\limits_{i=1}^\infty \sum\limits_{j=1}^\infty  \sum\limits_{k=1}^\infty v_i v_j (v_k + 3p_k) \delta_{|i+j-k|,0} \nonumber \\
    &+ 4 \sum\limits_{i=1}^\infty \sum\limits_{j=1}^\infty  \sum\limits_{k=1}^\infty v_i v_j (v_k + 3p_k) \delta_{|i-j+k|,0} \nonumber \\
    &+ 2 \sum\limits_{i=1}^\infty \sum\limits_{j=1}^\infty  \sum\limits_{k=1}^\infty  \sum\limits_{l=1}^\infty v_i v_j v_k p_l \delta_{|-i-j-k+l|,0} \nonumber \\
    &+ 6 \sum\limits_{i=1}^\infty \sum\limits_{j=1}^\infty  \sum\limits_{k=1}^\infty  \sum\limits_{l=1}^\infty v_i v_j v_k p_l \delta_{|-i+j+k+l|,0} \nonumber \\
    &+ 6 \sum\limits_{i=1}^\infty \sum\limits_{j=1}^\infty  \sum\limits_{k=1}^\infty  \sum\limits_{l=1}^\infty v_i v_j v_k p_l \delta_{|-i-j+k+l|,0} \nonumber 
\end{align}
and
\begin{align}
    C^B_n &= 6 v_n + 2 p_n + 6 \sum\limits_{i=1}^\infty \sum\limits_{j=1}^\infty v_i (v_j + p_j) \delta_{i+j,n} \\
    &+ 6 \sum\limits_{i=1}^\infty \sum\limits_{j=1}^\infty v_i (v_j + p_j) \delta_{|i-j|,n} \nonumber \\
    &+ 2 \sum\limits_{i=1}^\infty \sum\limits_{j=1}^\infty  \sum\limits_{k=1}^\infty v_i v_j (v_k + 3p_k) \delta_{i+j+k,n} \nonumber \\
    &+ 2 \sum\limits_{i=1}^\infty \sum\limits_{j=1}^\infty  \sum\limits_{k=1}^\infty v_i v_j (v_k + 3p_k) \delta_{|i+j-k|,n} \nonumber \\
    &+ 4 \sum\limits_{i=1}^\infty \sum\limits_{j=1}^\infty  \sum\limits_{k=1}^\infty v_i v_j (v_k + 3p_k) \delta_{|i-j+k|,n} \nonumber \\
    &+ 2 \sum\limits_{i=1}^\infty \sum\limits_{j=1}^\infty  \sum\limits_{k=1}^\infty  \sum\limits_{l=1}^\infty v_i v_j v_k p_l \delta_{i+j+k+l,n} \nonumber \\
    &+ 2 \sum\limits_{i=1}^\infty \sum\limits_{j=1}^\infty  \sum\limits_{k=1}^\infty  \sum\limits_{l=1}^\infty v_i v_j v_k p_l \delta_{|-i-j-k+l|,n} \nonumber \\
    &+ 6 \sum\limits_{i=1}^\infty \sum\limits_{j=1}^\infty  \sum\limits_{k=1}^\infty  \sum\limits_{l=1}^\infty v_i v_j v_k p_l \delta_{|-i+j+k+l|,n} \nonumber \\
    &+ 6 \sum\limits_{i=1}^\infty \sum\limits_{j=1}^\infty  \sum\limits_{k=1}^\infty  \sum\limits_{l=1}^\infty v_i v_j v_k p_l \delta_{|-i-j+k+l|,n} \nonumber 
\end{align}
We point out that this derivation trivially returns the results from Ref. \cite{Kolb:2004gi} when shadowing vanishes, i.e. for $p_n \rightarrow 0$. We will therefore not repeat in the following sections the discussion held in \cite{Kolb:2004gi}, which explained e.g. how a hadron elliptic flow can be generated without partonic elliptic flow, but purely from partonic directed flow ($v^M_2 \sim (v^q_1)^2$). We will instead focus on describing how different shadowing coefficients generate measureable hadron flow signals, without the respective signal on the partonic level or composite level before shadowing.


\subsection{Discussion}
We can expand the series expansion shown before, keeping only terms up to 4th order of each individual coefficient. We thus get the following expressions for the meson flow coefficients. The respective expressions for the baryons are shown and discussed in the appendix for brevity. 

The coefficients of the unshadowed meson flow are given by $\mathcal{V}^M_n = C^M_n / (2 C^M_0)$. We start the discussion with the directed flow $\mathcal{V}^M_1$ which is expressed as
\begin{align}
    &\mathcal{V}^M_1 = \frac{1}{C^M_0} [ 2 v_1 + 2 v_1 v_2 + 2 v_2 v_3 + 2 v_3 v_4 \\ 
    &+ p_1^M \left( 1 + 2 v_2 + 3 v_1^2 + 2 v_2^2 + 2 v_3^2 + 2 v_4^2 + 2 v_1 v_3 + 2 v_2 v_4 \right) \nonumber \\
    &+ p_2^M \left( 2 v_1 + 2 v_3 + 4 v_1 v_2 + 2 v_2 v_3 + 2 v_1 v_4 + 2 v_3 v_4 \right) \nonumber \\ 
    &+ p_3^M \left( 2 v_2 + 2 v_4 + v_1^2 + v_2^2 + 4 v_1 v_3 + 2 v_2 v_4 \right) \nonumber \\ 
    &+ p_4^M \left( 2 v_3 + 2 v_1 v_2 + 2 v_2 v_3 + 4 v_1 v_4 \right) ] \nonumber 
\end{align}
where one can clearly see that the standard result $\mathcal{V}^M_1 = 2 v_1$ for constituent quark number scaling is restored. However, the leading order expression now also has a directed shadowing component, i.e. $\mathcal{V}^M_1 = 2v_1 + p^M_1$. This might explain the increasing split between $K^+$ and $K^-$ directed flow\footnote{Of course other effects such as splitting via e.g. the electromagnetic interaction \cite{STAR:2023jdd} or the chiral magnetic wave \cite{STAR:2022zpv} influence this.} seen by STAR in \cite{STAR:2025twg} and preliminarily reported by HADES \cite{Orlinski:2025ozw}. The $K^+$ has a small cross section with nuclear matter and cannot form resonances with baryons, whereas the $K^-$ has a large cross section on nuclear matter and can form resonances. The effect becomes more enhanced at forward/backward rapidity. Next, the elliptic flow $\mathcal{V}^M_2$ reads
\begin{align}
    &\mathcal{V}^M_2 = \frac{1}{C^M_0} [ 2 v_2 + v_1^2 + 2 v_1 v_3 + 2 v_2 v_4 \\
    &+ p_1^M \left( 2 v_1 + 2 v_3 + 4 v_1 v_2 + 2 v_2 v_3 + 2 v_1 v_4 + 2 v_3 v_4 \right) \nonumber \\ 
    &+ p_2^M \left( 1 + 2 v_4 + 2 v_1^2 + 3 v_2^2 + 2 v_3^2 + 2 v_4^2 + 2 v_1 v_3 \right) \nonumber \\
    &+ p_3^M \left( 2 v_1 + 2 v_1 v_2 + 4 v_2 v_3 + 2 v_1 v_4 + 2 v_3 v_4 \right) \nonumber \\ 
    &+ p_4^M \left( 2 v_2 + v_1^2 + v_3^2 + 2 v_1 v_3 + 4 v_2 v_4 \right) ] \nonumber
\end{align}
Again, we recover the standard result and get a leading order shadowing correction $\mathcal{V}^M_2 = 2v_2 + p^M_2$. Therefore, even an in-plane expanding fireball might yield a negative final state elliptic flow, when there is sufficient elliptic shadowing contribution. This exactly is what makes the proton elliptic flow negative at low collision energies as seen by many experiments (cf. compilation in \cite{HADES:2022osk}). Interestingly, a finite elliptic flow at forward/backward rapidity can also be generated from the competition of the directed flow component and directed shadowing $p_1^M v_1$, e.g. explaining pion elliptic flow \cite{STAR:2021yiu,Prozorov:2023lyv}. The triangular flow coefficient $\mathcal{V}^M_3$ becomes
\begin{align}
    &\mathcal{V}^M_3 = \frac{1}{C^M_0} [2 v_3 + 2 v_1 v_2 + 2 v_1 v_4 \\
    &+ p_1^M \left( 2 v_2 + 2 v_4 + v_1^2 + v_2^2 + 4 v_1 v_3 + 2 v_2 v_4 \right) \nonumber \\
    &+ p_2^M \left( 2 v_1 + 2 v_1 v_2 + 4 v_2 v_3 + 2 v_1 v_4 + 2 v_3 v_4 \right) \nonumber \\
    &+ p_3^M \left( 1 + 2 v_1^2 + 2 v_2^2 + 3 v_3^2 + 2 v_4^2 + 2 v_2 v_4 \right) \nonumber \\
    &+ p_4^M \left( 2 v_1 + 2 v_1 v_2 + 2 v_2 v_3 + 4 v_3 v_4 \right) ] \nonumber
\end{align}
The leading order is, again, corrected by a triangular shadowing component, i.e. $\mathcal{V}^M_3 = 2v_3 + p^M_3$. Investigating the triangular flow is particularly interesting, because it is often related to scaling relations, where $\mathcal{V}^h_3 \sim \mathcal{V}^h_1 \mathcal{V}^h_2$, and due to the lacking triangular symmetry of the collision system. Without shadowing, a triangular meson flow coefficient could have only been generated from partonic triangular flow or from the product of partonic directed and elliptic flow. When one includes shadowing, there are now many more ways to generate a meson triangular flow, without requiring higher order symmetries in the system. A finite $\mathcal{V}^M_3$ can hence be generated by e.g. the following terms: $p^M_1 v_2$, $p^M_1 v_1^2$, $p_1^M v_2^2$, $p^M_2 v_1$, $p^M_2 v_1 v_2$. This also underlines the commonly attributed geometric origin of triangular flow \cite{HADES:2022osk,STAR:2023duf}, for which it is here quantitatively clear how the geometry generates a triangular flow. For the quadrangular flow we then get
\begin{align}
    &\mathcal{V}^M_4 = \frac{1}{C^M_0} [ 2 v_4 + v_2^2 + 2 v_1 v_3 \\
    &+ p_1^M \left( 2 v_3 + 2 v_1 v_2 + 2 v_2 v_3 + 4 v_1 v_4 \right) \nonumber \\
    &+ p_2^M \left( 2 v_2 + v_1^2 + v_3^2 + 2 v_1 v_3 + 4 v_2 v_4 \right) \nonumber \\
    &+ p_3^M \left( 2 v_1 + 2 v_1 v_2 + 2 v_2 v_3 + 4 v_3 v_4 \right) \nonumber \\
    &+ p_4^M \left( 1 + 2 v_1^2 + 2 v_2^2 + 2 v_3^2 + 3 v_4^2 \right) ] \nonumber
\end{align}
As expected, we again observe that $\mathcal{V}^M_4 = 2v_4 + p^M_4$ at the leading order. However, a quadrangular signal can also be generated by combinations of first and second order coefficients (which is not possible without shadowing, i.e. there is no term appearing proportional to $v_1 v_2$). These terms are: $p^M_1 v_1 v_2$, $p^M_2 v_2$, $p^M_2 v_1^2$. Finally, the normalization $V^M_0$ is 
\begin{align}
    &C^M_0 = 1 + 2 v_1^2 + 2 v_2^2 + 2 v_3^2 + 2 v_4^2 \\
    &+ p_1^M (4 v_1 + 4 v_1 v_2 + 4 v_2 v_3 + 4 v_3 v_4) \nonumber \\
    &+ p_2^M (4 v_2 + 2 v_1^2 + 4 v_1 v_3 + 4 v_2 v_4) \nonumber \\
    &+ p_3^M (4 v_3 + 4 v_1 v_2 + 4 v_1 v_4) \nonumber \\
    &+ p_4^M (4 v_4 + 2 v_2^2 + 4 v_1 v_3) \nonumber
\end{align}
where the leading shadowing coefficients provide corrections with equal orders of shadowing and parton flow, i.e. $p^M_n v_n$.

\begin{figure*} [t!hb]
    \centering
    \includegraphics[width=2\columnwidth]{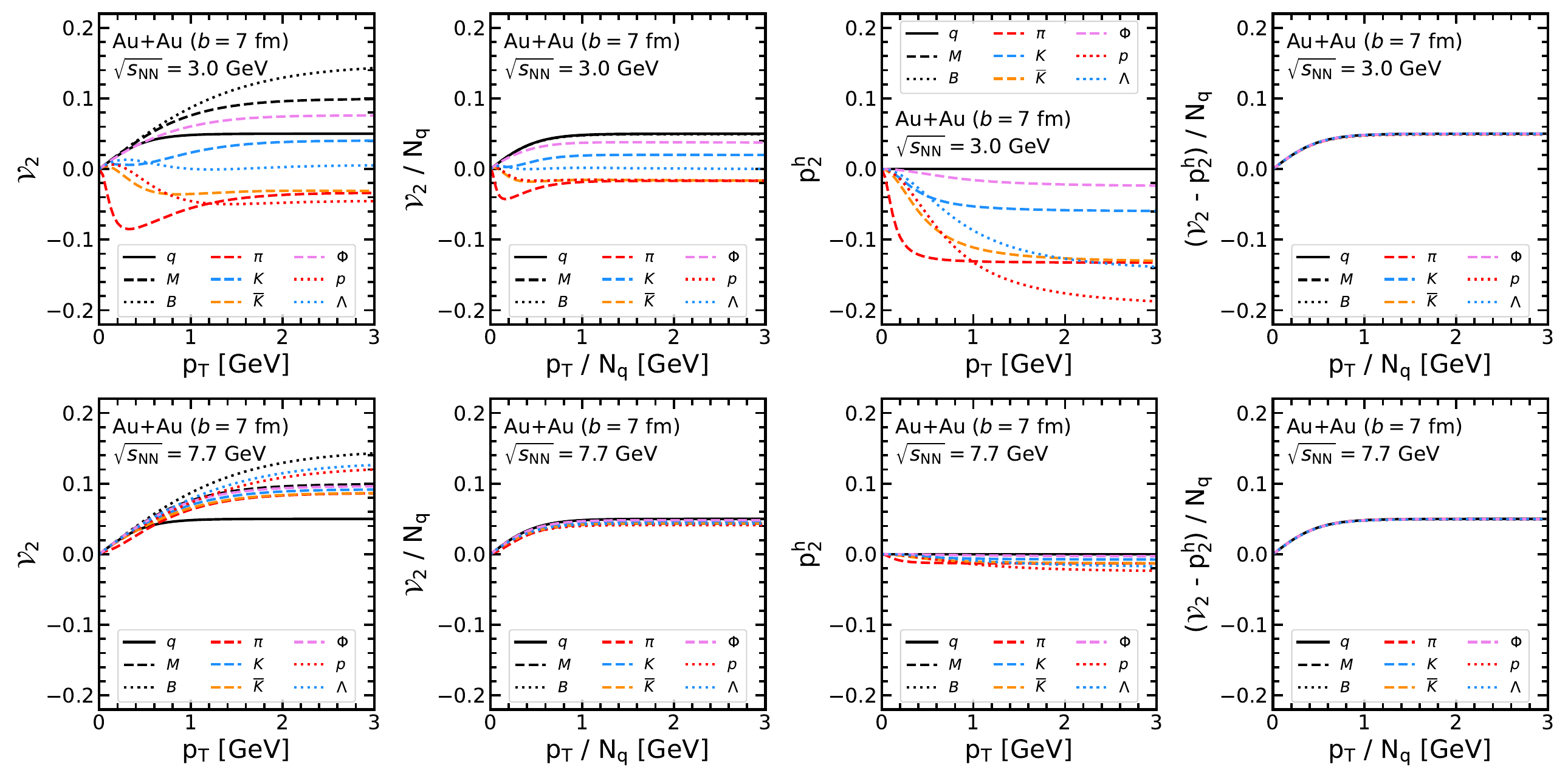}
    \caption{[Color online] The measureable elliptic flow $\mathcal{V}_2$ (first column), scaled elliptic flow $\mathcal{V}_2/N_q$ (second column), second order shadowing coefficient $p_2^h$ (third column) and scaled unshadowed elliptic flow $(\mathcal{V}_2 - p_2^h)/N_q$ (fourth column) of $\pi$ (dashed red line), $K$ (dashed blue line), $\overline{K}$ (dashed orange line), $\Phi$ (dashed pink line), $p$ (dotted red line) and $\Lambda$ (dotted blue line) in comparison to the idealized results without shadowing of partons (solid black line), mesons (dashed black line) and baryons (dotted black line) in peripheral Au+Au collisions ($b=7$ fm) at $\sqrt{s_\mathrm{NN}}=3.0$ GeV (upper row) and $\sqrt{s_\mathrm{NN}}=7.7$ GeV (lower row). The results were calculated assuming that hadrons are formed by quark coalescence and emitted isotropically at $t_\mathrm{overlap}$ at the origin at midrapidity $y=0$. The shadowing coefficients were calculated using the presented toy model consisting of a ballistic Glauber density profile and constant effective absorption cross sections for the hadrons.}
    \label{fig:toy_model}
\end{figure*}
\section{Toy model estimates and qualitative effect}
It will, obviously, be very valuable to find (semi-)analytic expressions for the shadowing coefficients $p^h_n$ that factor in the collision energy $\sqrt{s_\mathrm{NN}}$ (and therefore the passing time $t_\mathrm{pass}$ of the spectator) as well as the cross section. That way one could correct measured data for the shadowing contribution and investigate whether constituent quark number scaling is present in the hadron emitting source. In principle this task is very well suited for transport models or kinetic theory calculations, which factor in the relevant passing time, the energy and angle dependent cross section and resonance dynamics. 

However, in order to demonstrate the qualitative effect that shadowing has on a source emitting hadrons that obey constituent quark number scaling we will adopt a toy model that accounts for the effective shadowing. We will calculate the measureable elliptic flow $\mathcal{V}_2$ of several hadron species emitted at midrapidity $y=0$ (this will get rid of all odd coefficients due to symmetry) purely from a parametrized parton $v_2$ and a simple transverse absorption probability. For this we parametrize the parton $v_2$ in a commonly employed form as
\begin{align}
    v_2^q(p_\mathrm{T}^q) &= v_{2,\mathrm{max}}^q \tanh{\left(p_\mathrm{T}^q / \Lambda\right)},
\end{align}
where $v_2^q$ is the parton's elliptic flow, $v_{2,\mathrm{max}}^q$ is the saturation value of the elliptic flow, $p_\mathrm{T}^q$ is the parton's transverse momentum and $\Lambda$ is a scale. In coherence with previous calculations \cite{Dunlop:2011cf} we will set $v_{2,\mathrm{max}}^q = 0.05$ and $\Lambda = 0.5$ GeV. This fully determines the hadron's elliptic flow before shadowing. The shadowing coefficients are qualitatively estimated from a toy model. For this we employ a ballistic Glauber model as e.g. developed in \cite{Vovchenko:2014gda}. Therein the density
\begin{align}
    \rho(t,r) &= \frac{\gamma \rho_0}{1 + \exp \left( \frac{r(t) - R_0}{\sigma_r} \right)} \label{eq:WS} \\
    r(t) &= \sqrt{\left(x \mp \frac{b}{2}\right)^2 + y^2 + \gamma^2 \left(z \pm \frac{R_0}{\gamma} \mp \beta t\right)^2} \label{eq:r(t)}
\end{align}
is modeled by a smooth Woods-Saxon profile propagating ballistically, i.e. without baryon stopping. Herein, the nuclei are initialized with a given impact parameter $b$ and longitudinal offset of $R_0/\gamma$. Thus the density used to calculate the absorption probability $P_\mathrm{esc}$ depends only on the system parameters $\sqrt{s_\mathrm{NN}}$, $b$ and $A$ ($R_0$). This serves as a first order approximation for the density evolution which is required to calculate the shadowing coefficients from the escape probability $P_\mathrm{esc}(\phi)$ shown in eq. \eqref{eq:Pesc}. We further assume that all hadrons are produced by quark coalescence and are emitted from a point-like source at the origin at the time of full overlap of the two nuclei, i.e. at $t_\mathrm{overlap} = R / (\beta\gamma)$, and that the initial emission profile is isotropic. The transverse velocity of the hadrons is obviously given by $p_\mathrm{T}/m_\mathrm{T}$ at midrapidity\footnote{Even when all cross sections were one common constant for all hadrons, constituent quark number scaling would still be broken. This is because the hadrons will have a different velocity at equal transverse momentum and thus propagate faster or slower through the spectator.}. 

We then calculate the density evolution from eqs. \eqref{eq:WS}, \eqref{eq:r(t)} for a peripheral Au+Au collision at $b=7$ fm at $\sqrt{s_\mathrm{NN}}=3.0$ GeV and at $\sqrt{s_\mathrm{NN}}=7.7$ GeV, representing two distinct regimes. We calculate the measureable elliptic flow for mesons and baryons without shadowing and for the selected hadrons $\pi$, $K$, $\overline{K}$, $\Phi$, $p$ and $\Lambda$. We adopt constant cross sections that resemble the effective absorption of these hadrons on the bypassing spectator. Those are: $\sigma^\mathrm{eff}_{\pi N} = 50$ mb, $\sigma^\mathrm{eff}_{K N} = 12$ mb, $\sigma^\mathrm{eff}_{\overline{K} N} = 50$ mb, $\sigma^\mathrm{eff}_{\phi N} = 5$ mb, $\sigma^\mathrm{eff}_{p N} = 40$ mb and $\sigma^\mathrm{eff}_{\Lambda N} = 30$ mb. The values of the $\pi$ and $\overline{K}$ are chosen smaller than their actual values, because both mesons are subject to reoccurring resonance excitation and decay via $\Delta \leftrightarrow \pi N$ and $\Lambda^\ast \leftrightarrow \overline{K} N$, which will repopulate their yields strongly at low energies. 

In Fig. \ref{fig:toy_model} we show the results (see legend for color identification). The upper and lower rows depict the results at $\sqrt{s_\mathrm{NN}}=3.0$ GeV and at $\sqrt{s_\mathrm{NN}}=7.7$ GeV, respectively. The first column shows the measureable elliptic flow $\mathcal{V}_2$ of the coalescing source after the hadrons experienced shadowing. One can clearly observe that the influence of shadowing is much enhanced at the lower collision energy due to longer passing time of the spectator. The $p$ and $\Lambda$ elliptic flow values differ from the idealized baryon elliptic flow without shadowing, while for the mesons the difference from the idealized scenario is even stronger. The second column shows the elliptic flow naively scaled by the number of constituent quarks $\mathcal{V}_2/N_q$ as a function of the scaled transverse momentum. While at the higher collision energy (with faster decoupling spectator) the scaling is nearly perfect, it is clearly broken at the lower collision energy. This matches the qualitative observations by STAR \cite{STAR:2021yiu,STAR:2025owm}. The third column shows the second order shadowing coefficient $p_2^h$ of all hadron species. Here, one can observe a clear hierarchy of the hadrons based on their interaction cross section with the medium. Also due to the different masses of the hadrons, the shadowing contribution saturates at different transverse momenta. Finally, the fourth column depicts the corrected (``unshadowed'') elliptic flow scaled by the number of constituent quarks, i.e. $(\mathcal{V}_2 - p_2^h) / N_q \equiv V_2 / N_q$ as a function of the scaled transverse momentum. As expected (and due to construction in the toy model) the shadowing coefficient $p_2^h$ cancels the contribution from spectator shadowing and yields the elliptic flow of the unshadowed hadrons, which then scaled with the number of constituent quarks. 

We note here explicitly, that this toy model is not sophisticated enough to appropriately unshadow measured experimental data and answer whether constituent quark number scaling is present in the STAR-FXT and FAIR energy regimes, but it qualitatively demonstrates how such a correction can be calculated. In principle the task to calculate all $p^h_n$ coefficients is very well suited for hadronic transport models, as they include energy (and angular) dependent cross sections $\sigma(s,t)$ as well as resonance dynamics and their subsequent chain reactions that e.g. strongly populate the pion number via $\Delta \rightarrow \pi N$ or $\rho \rightarrow \pi \pi$ reactions and they also include transverse expansion effects and stopping. We will thus examine the absorption coefficients $p_n^h$ in an upcoming study \cite{Reichert:tbp} using a hadronic transport model with the aim to provide an effective correction for shadowing in the STAR-FXT and FAIR energy regimes.

\section{Conclusion and outlook}
In this article we have studied constituent quark number scaling under the influence of spectator shadowing. We have derived analytic expressions that connect the measureable flow coefficients to the flow coefficients of the hadron emitting source and the Fourier coefficients of the transverse escape probability of hadrons propagating through the spectator. We observe, that in the presence of drastic shadowing, one cannot expect that the measured flow coefficients fulfill constituent quark number scaling directly, even if quark coalescence were the driving factor of hadron formation. Instead, the unshadowed flow coefficients now fulfill constituent quark number scaling. Thus one obtains the leading order scaling relations.
\begin{align}
    \mathcal{V}^M_n(p_\mathrm{T}) - p^M_n(p_\mathrm{T}) = 2 v_n(p_\mathrm{T} / 2) \\
    \mathcal{V}^B_n(p_\mathrm{T}) - p^B_n(p_\mathrm{T}) = 3 v_n(p_\mathrm{T} / 3)
\end{align}
The naively assumed scaling of the measured $\mathcal{V}^h_n = N_q v_n$ (without shadowing) thus first breaks for hadrons with a large cross sections and last for hadrons with a small cross section. The scaling breaks continuously when decreasing the collision energy, i.e. the shadowing $p_n$ coefficients decrease smoothly with the spectators' passing time. The derived equations provide a natural explanation for the generation of higher order flow components without invoking higher order flow of the partonic source and it thus quantifies how the geometry shapes the measureable flow. We have then adopted a toy model using a ballistic Glauber model to calculate the qualitative impact of shadowing on a coalescing source. The model calculations match the recent STAR measurement \cite{STAR:2021yiu,STAR:2025owm} qualitatively. The measured results do therefore neither rule out the presence of partonic collectivity, nor do they prove it, if taken by face value. 

The results are also relevant to understand $\phi$ flow, as preliminarily reported by STAR, and under theoretical investigation \cite{Steinheimer:2025mho}. The $\phi$ might serve as a non-shadowed baseline due to its small cross section (this similarly applies for di-leptons, although they provide a more space-time integrated result).

We have thus provided, for the first time, a qualitative and quantitative understanding of how parton coalescence is affected by shadowing and how it shows up in measured data. The derived equations can directly be employed to correct measured data and investigate whether constituent quark number scaling has been present at the hadron emitting source. 

\section*{Acknowledgements}
We thank Agnieszka Sorensen, Paul Sorensen, Richard Seto, Volodymyr Vovchenko and Volker Koch for fruitful discussions about NCQ scaling during the INT workshop ``The QCD Critical Point: Are We There Yet?'' under event code INT-25-3a that have inspired this study. 
We thank the Institute for Nuclear Theory at the University of Washington for its kind hospitality and stimulating research environment. This research was supported in part by the INT's U.S. Department of Energy grant No. DE-FG02-00ER41132.
T.R. thanks Steffen Bass for the kind hospitality at Duke University. 
T.R. gratefully acknowledges financial support by the Fulbright U.S. Scholar Program, which is sponsored by the U.S. Department of State and the German-American Fulbright Commission. This article’s contents are solely the responsibility of the author and do not necessarily represent the official views of the Fulbright Program, the Government of the United States, or the German-American Fulbright Commission. 
T.R. gratefully acknowledges support from The Branco Weiss Fellowship - Society in Science, administered by the ETH Z\"urich. IK acknowledges support by the Czech Science Foundation under project No.~25-16877S.

\appendix

\section{Baryon coefficients}
In the following we show the unshadowed baryon flow coefficients $\mathcal{V}^B_n = C^B_n / (2 C^B_0)$ up to fourth order of each individual coefficient and provide a discussion of the leading influence of the shadowing coefficients $p_n$.
\begin{widetext}
The directed flow of baryons $\mathcal{V}^B_1$ is given by 
\begin{align}
    &\mathcal{V}^B_1 = \frac{1}{C^B_0} \biggl[ 3 v_1 + 6 v_1 v_2 + 6 v_2 v_3 + 6 v_3 v_4 + 3 v_1^3 + 6 v_1 v_2^2 + 3 v_1^2 v_3 + 3 v_2^2 v_3 + 6 v_1 v_3^2 + 6 v_1 v_4^2 + 6 v_1 v_2 v_4 + 6 v_2 v_3 v_4 \\
    &+ p_1^B (1 + 3 v_2 + 9 v_1^2 + 6 v_2^2 + 6 v_3^2 + 6 v_4^2 + 6 v_1 v_3 + 6 v_2 v_4 + 3 v_2^3 + 12 v_1^2 v_2 + 6 v_2 v_3^2 + 3 v_1^2 v_4 + 6 v_2^2 v_4 \nonumber \\
    &\qquad\; + 3 v_3^2 v_4 + 6 v_2 v_4^2 + 18 v_1 v_2 v_3 + 18 v_1 v_3 v_4) \nonumber \\
    &+ p_2^B (3 v_1 + 3 v_3 + 4 v_1^3 + 3 v_3^3 + 12 v_1 v_2 + 9 v_1 v_2^2 + 9 v_1^2 v_3 + 6 v_2 v_3 + 9 v_2^2 v_3 + 6 v_1 v_3^2 + 6 v_1 v_4 + 6 v_3 v_4 + 12 v_1 v_2 v_4 \nonumber \\
    &\qquad\; + 12 v_2 v_3 v_4 + 6 v_1 v_4^2 + 6 v_3 v_4^2) \nonumber \\
    &+ p_3^B (3 v_2 + 3 v_4 + 3 v_1^2 + 3 v_2^2 + 12 v_1 v_3 + 6 v_2 v_4 + 3 v_2^3 + 3 v_4^3 + 9 v_1^2 v_2 + 9 v_2 v_3^2 + 9 v_1^2 v_4 + 6 v_2^2 v_4 + 9 v_3^2 v_4 \nonumber \\
    &\qquad\; + 6 v_2 v_4^2 + 12 v_1 v_2 v_3 + 6 v_1 v_3 v_4) \nonumber \\
    &+ p_4^B (3 v_3 + 6 v_1 v_2 + 6 v_2 v_3 + 12 v_1 v_4 + v_1^3 + 3 v_3^3 + 6 v_1 v_2^2 + 9 v_1^2 v_3 + 6 v_2^2 v_3 + 3 v_1 v_3^2  + 9 v_3 v_4^2 + 12 v_1 v_2 v_4 + 12 v_2 v_3 v_4) \biggr] \nonumber
\end{align}
which recovers the standard result as well. Here, the leading order correction is $\mathcal{V}^B_1 = 3v_1 + p^B_1$. In similar way as the directed meson flow could explain the split in Kaon flow, the baryon directed flow explains the split between baryon and anti-baryon flow ($p, \Lambda$ vs. $\Bar{p}, \Bar{\Lambda}$). Next, the elliptic flow of baryons $\mathcal{V}^B_2$ is given by 
\begin{align}
    &\mathcal{V}^B_2 = \frac{1}{C^B_0} \biggl[ 3 v_2 + 3 v_1^2 + 6 v_1 v_3 + 6 v_2 v_4 + 3 v_2^3 + 6 v_1^2 v_2 + 6 v_2 v_3^2 + 3 v_1^2 v_4 + 3 v_3^2 v_4 + 6 v_2 v_4^2 + 6 v_1 v_2 v_3 + 6 v_1 v_3 v_4 \\
    &+ p_1^B (3 v_1 + 3 v_3 + 12 v_1 v_2 + 6 v_1 v_4 + 6 v_3 v_4 + 6 v_2 v_3 + 4 v_1^3 + 3 v_3^3 + 9 v_1 v_2^2 + 9 v_1^2 v_3 + 9 v_2^2 v_3 + 6 v_1 v_3^2 + 6 v_1 v_4^2 + 6 v_3 v_4^2 \nonumber \\
    &\qquad\; + 12 v_1 v_2 v_4 + 12 v_2 v_3 v_4) \nonumber \\
    &+ p_2^B (1 + 3 v_4 + 6 v_1^2 + 9 v_2^2 + 6 v_3^2 + 6 v_4^2 + 6 v_1 v_3 + 3 v_4^3 + 9 v_1^2 v_2 + 3 v_2 v_3^2 + 6 v_1^2 v_4 + 12 v_2^2 v_4 + 6 v_3^2 v_4 \nonumber \\
    &\qquad\; + 18 v_1 v_2 v_3 + 12 v_1 v_3 v_4) \nonumber \\
    &+ p_3^B (3 v_1 + 6 v_1 v_2 + 12 v_2 v_3 + 6 v_1 v_4 + 6 v_3 v_4 + 3 v_1^3 + 9 v_1 v_2^2 + 6 v_1^2 v_3 + 3 v_2^2 v_3 + 9 v_1 v_3^2 + 6 v_1 v_4^2 + 3 v_3 v_4^2 \nonumber \\
    &\qquad\; + 12 v_1 v_2 v_4 + 12 v_2 v_3 v_4) \nonumber \\
    &+ p_4^B (3 v_2 + 3 v_1^2 + 3 v_3^2 + 6 v_1 v_3 + 12 v_2 v_4 + 4 v_2^3 + 6 v_1^2 v_2 + 6 v_2 v_3^2 + 6 v_1^2 v_4 + 3 v_3^2 v_4 + 9 v_2 v_4^2 + 12 v_1 v_2 v_3 + 12 v_1 v_3 v_4) \biggr] \nonumber
\end{align}
where, again, the standard results in the literature are reproduced if no shadowing is present. The leading order correction is $\mathcal{V}^B_2 = 3v_2 + p^B_2$. This is clearly the main contribution that leads to negative elliptic flow of protons (baryons in general) at RHIC-FXT and GSI collision energies. The triangular component $\mathcal{V}^B_3$ becomes
\begin{align}
    &\mathcal{V}^B_3 = \frac{1}{C^B_0} \biggl[ 3 v_3 + 6 v_1 v_2 + 6 v_1 v_4 + v_1^3 + 3 v_3^3 + 3 v_1 v_2^2 + 6 v_1^2 v_3 + 6 v_2^2 v_3 + 6 v_3 v_4^2 + 6 v_1 v_2 v_4 + 6 v_2 v_3 v_4 \\
    &+ p_1^B (3 v_2 + 3 v_4 + 3 v_1^2 + 3 v_2^2 + 12 v_1 v_3 + 6 v_2 v_4 + 3 v_2^3 + 3 v_4^3 + 9 v_1^2 v_2 + 9 v_2 v_3^2 + 9 v_1^2 v_4 + 6 v_2^2 v_4 + 9 v_3^2 v_4 + 6 v_2 v_4^2 \nonumber \\
    &\qquad\; + 12 v_1 v_2 v_3 + 6 v_1 v_3 v_4) \nonumber \\
    &+ p_2^B (3 v_1 + 6 v_1 v_2 + 12 v_2 v_3 + 6 v_1 v_4 + 6 v_3 v_4 + 3 v_1^3 + 9 v_1 v_2^2 + 6 v_1^2 v_3 + 3 v_2^2 v_3 + 9 v_1 v_3^2 + 6 v_1 v_4^2 + 3 v_3 v_4^2 \nonumber \\
    &\qquad\; + 12 v_1 v_2 v_4 + 12 v_2 v_3 v_4) \nonumber \\
    &+ p_3^B (1 + 6 v_1^2 + 6 v_2^2 + 9 v_3^2 + 6 v_4^2 + 6 v_2 v_4 + v_2^3 + 6 v_1^2 v_2 + 3 v_1^2 v_4 + 6 v_2^2 v_4 + 3 v_2 v_4^2 + 18 v_1 v_2 v_3 + 18 v_1 v_3 v_4) \nonumber \\
    &+ p_4^B (3 v_1 + 6 v_1 v_2 + 6 v_2 v_3 + 12 v_3 v_4 + 3 v_1^3 + 6 v_1 v_2^2 + 3 v_1^2 v_3 + 6 v_2^2 v_3 + 9 v_1 v_3^2 + 9 v_1 v_4^2 + 12 v_1 v_2 v_4 + 6 v_2 v_3 v_4) \biggr] \nonumber
\end{align}
also reproducing the standard result and having the leading order correction $\mathcal{V}^B_3 = 3v_3 + p^B_3$. In a similar way that was already present for the mesons, a measureable triangular flow can be purely generated from first and second order combinations of partonic flow and shadowing coefficients. These are $p^B_1 v_2$, $p^B_1 v_1^2$, $p^B_1 v_2^2$, $p^B_1 v_2^3$ and $p^B_1 v_1^2 v_2$ for the directed shadowing and $p^B_2 v_1$, $p^B_2 v_1^3$, $p^B_2 v_1 v_2$ and $p^B_2 v_1 v_2^2$ for the elliptic shadowing. Especially, the respective first terms are interesting ($p^B_1 v_2$ and $p^B_2 v_1$), because they explain a finite triangular flow by direct combinations of partonic flow and shadowing. Finally, the quadrangular barony flow $\mathcal{V}^B_4$ is
\begin{align}
    &\mathcal{V}^B_4 = \frac{1}{C^B_0} \biggl[ 3 v_4 + 3 v_2^2 + 6 v_1 v_3 + 3 v_4^3 + 3 v_1^2 v_2 + 3 v_2 v_3^2 + 6 v_1^2 v_4 + 6 v_2^2 v_4 + 6 v_3^2 v_4 + 6 v_1 v_2 v_3 \\
    &+ p_1^B (3 v_3 + 6 v_1 v_2 + 6 v_2 v_3 + 12 v_1 v_4 + v_1^3 + 3 v_3^3 + 6 v_1 v_2^2 + 9 v_1^2 v_3 + 6 v_2^2 v_3 + 3 v_1 v_3^2 + 9 v_3 v_4^2 + 12 v_1 v_2 v_4 + 12 v_2 v_3 v_4) \nonumber \\
    &+ p_2^B (3 v_2 + 3 v_1^2 + 3 v_3^2 + 6 v_1 v_3 + 12 v_2 v_4 + 4 v_2^3 + 6 v_1^2 v_2 + 6 v_2 v_3^2 + 6 v_1^2 v_4 + 3 v_3^2 v_4 + 9 v_2 v_4^2 + 12 v_1 v_2 v_3 + 12 v_1 v_3 v_4) \nonumber \\
    &+ p_3^B (3 v_1 + 6 v_1 v_2 + 6 v_2 v_3 + 12 v_3 v_4 + 3 v_1^3 + 6 v_1 v_2^2 + 3 v_1^2 v_3 + 6 v_2^2 v_3 + 9 v_1 v_3^2 + 9 v_1 v_4^2 + 12 v_1 v_2 v_4 + 6 v_2 v_3 v_4) \nonumber \\
    &+ p_4^B (1 + 6 v_1^2 + 6 v_2^2 + 6 v_3^2 + 9 v_4^2 + 6 v_1^2 v_2 + 3 v_2 v_3^2 + 9 v_2^2 v_4 + 12 v_1 v_2 v_3 + 18 v_1 v_3 v_4) \biggr] \nonumber
\end{align}
which also reproduces the known result and is linear in its leading order $\mathcal{V}^B_4 = 3v_4 + p^B_4$. Again, one can generate a finite quadrangular flow from the following combinations of first and second order partonic and shadowing coefficients: $p^B_1 v_1^3$, $p^B_1 v_1 v_2$, $p^B_1 v_1 v_2^2$, $p^B_2 v_2$, $p^B_2 v_1^2$, $p^B_2 v_2^3$ and $p^B_2 v_1^2 v_2$. The normalization is
\begin{align}
    &C^B_0 = 1 + 6 v_1^2 + 6 v_2^2 + 6 v_3^2 + 6 v_4^2 + 6 v_1^2 v_2 + 6 v_2^2 v_4 + 12 v_1 v_2 v_3 + 12 v_1 v_3 v_4 \\
    &+ p_1^B (6 v_1 + 12 v_1 v_2 + 12 v_2 v_3 + 12 v_3 v_4 + 6 v_1^3 + 12 v_1 v_2^2 + 6 v_1^2 v_3 + 6 v_2^2 v_3 + 12 v_1 v_3^2 + 12 v_1 v_4^2 + 12 v_1 v_2 v_4 + 12 v_2 v_3 v_4) \nonumber \\
    &+ p_2^B (6 v_2 + 6 v_1^2 + 12 v_1 v_3 + 12 v_2 v_4 + 6 v_2^3 + 12 v_1^2 v_2 + 12 v_2 v_3^2 + 6 v_1^2 v_4 + 6 v_3^2 v_4 + 12 v_2 v_4^2 + 12 v_1 v_2 v_3 + 12 v_1 v_3 v_4) \nonumber \\
    &+ p_3^B (6 v_3 + 12 v_1 v_2 + 12 v_1 v_4 + 2 v_1^3 + 6 v_3^3 + 6 v_1 v_2^2 + 12 v_1^2 v_3 + 12 v_2^2 v_3 + 12 v_3 v_4^2 + 12 v_1 v_2 v_4 + 12 v_2 v_3 v_4) \nonumber \\
    &+ p_4^B (6 v_4 + 6 v_2^2 + 12 v_1 v_3 + 6 v_4^3 + 6 v_1^2 v_2 + 6 v_2 v_3^2 + 12 v_1^2 v_4 + 12 v_2^2 v_4 + 12 v_3^2 v_4 + 12 v_1 v_2 v_3) \nonumber
\end{align}
where the leading shadowing coefficients provide corrections with equal orders of shadowing and parton flow, i.e. $p^B_n v_n$.
\end{widetext}

\section{Derivation}
Here we provide the main steps for the derivation of the shadowing contribution to the measureable flow signal. We shortly recapitulate that the measureable flow signal of composite mesons (baryons) that were formed by quark coalescence is given by $\mathcal{F}_M(\phi) \propto f_q^2(\phi) P_\mathrm{esc}(\phi)$ ($\mathcal{F}_B(\phi) \propto f_q^3(\phi) P_\mathrm{esc}(\phi)$). We can formally express all appearing functions as Fourier series with the coefficients $v_i$ (quarks), $p_i$ (shadowing), $\mathcal{V}_n$ for the mesons and baryons and we adopt the typical notation of the community, i.e. that the Fourier series goes as $1 + 2 \sum_{i=1}^\infty v_i \cos(i\phi)$. One then multiplies the Fourier series of the quarks and the shadowing contribution and makes use of the trigonometric relations for the product of two, three or four cosines. This yields
\begin{align}
    \mathcal{F}_M(\phi) &\propto 1 + 2 \sum\limits_{i=1}^\infty (2v_i + p_i) \cos(i)\phi \\
    &+ 2 \sum\limits_{i=1}^\infty \sum\limits_{j=1}^\infty v_i (v_j + 2p_j) \cos(i+j)\phi \nonumber \\
    &+ 2 \sum\limits_{i=1}^\infty \sum\limits_{j=1}^\infty v_i (v_j + 2p_j) \cos(-i+j)\phi \nonumber \\
    &+ 4 \sum\limits_{i=1}^\infty \sum\limits_{j=1}^\infty \sum\limits_{k=1}^\infty v_i v_j p_k \cos(-i+j+k)\phi \nonumber \\
    &+ 2 \sum\limits_{i=1}^\infty \sum\limits_{j=1}^\infty \sum\limits_{k=1}^\infty v_i v_j p_k \cos(-i-j+k)\phi \nonumber \\
    &+ 2 \sum\limits_{i=1}^\infty \sum\limits_{j=1}^\infty \sum\limits_{k=1}^\infty v_i v_j p_k \cos(i+j+k)\phi \nonumber 
\end{align}
for the mesons and
\begin{align}
    \mathcal{F}_B(\phi) &\propto 1 + 2 \sum\limits_{i=1}^\infty (3v_i + p_i) \cos(i)\phi \\
    + 6 \sum\limits_{i=1}^\infty  &\sum\limits_{j=1}^\infty v_i (v_j + p_j) \cos(i+j)\phi \nonumber \\
    + 6 \sum\limits_{i=1}^\infty  &\sum\limits_{j=1}^\infty v_i (v_j + p_j) \cos(i-j)\phi \nonumber \\
    + 2 \sum\limits_{i=1}^\infty  &\sum\limits_{j=1}^\infty  \sum\limits_{k=1}^\infty v_i v_j (v_k + 3p_k) \cos(i+j+k)\phi \nonumber \\
    + 2 \sum\limits_{i=1}^\infty  &\sum\limits_{j=1}^\infty  \sum\limits_{k=1}^\infty v_i v_j (v_k + 3p_k) \cos(i+j-k)\phi \nonumber \\
    + 4 \sum\limits_{i=1}^\infty  &\sum\limits_{j=1}^\infty  \sum\limits_{k=1}^\infty v_i v_j (v_k + 3p_k) \cos(i-j+k)\phi \nonumber \\
    + 2 \sum\limits_{i=1}^\infty  &\sum\limits_{j=1}^\infty  \sum\limits_{k=1}^\infty  \sum\limits_{l=1}^\infty v_i v_j v_k p_l \cos(i+j+k+l)\phi \nonumber \\
    + 2 \sum\limits_{i=1}^\infty  &\sum\limits_{j=1}^\infty  \sum\limits_{k=1}^\infty  \sum\limits_{l=1}^\infty v_i v_j v_k p_l \cos(-i-j-k+l)\phi \nonumber \\
    + 6 \sum\limits_{i=1}^\infty  &\sum\limits_{j=1}^\infty  \sum\limits_{k=1}^\infty  \sum\limits_{l=1}^\infty v_i v_j v_k p_l \cos(-i+j+k+l)\phi \nonumber \\
    + 6 \sum\limits_{i=1}^\infty  &\sum\limits_{j=1}^\infty  \sum\limits_{k=1}^\infty  \sum\limits_{l=1}^\infty v_i v_j v_k p_l \cos(-i-j+k+l)\phi \nonumber 
\end{align}
for the baryons. Now one can do a comparison of coefficients with the Fourier series of the measureable hadron flow, i.e. $\mathcal{F}_{M,B}(\phi) = 1 + 2 \sum_{n=1}^\infty \mathcal{V}^{M,B}_n \cos(n \phi)$. From this it is already evident that the measureable flow coefficients are
\begin{align}
    \mathcal{V}^{M,B}_n &= \frac{C^{M,B}_n}{2 C^{M,B}_0},
\end{align}
where the coefficients $C^{M,B}_0$ and $C^{M,B}_n$ collect all terms from the aforementioned expressions that yield the modularity $\cos(0\phi)$ and $\cos(n\phi)$, respectively. Some of the infinite sums are automatically truncated by this procedure, because the $v_i$ and $p_i$ are not defined for zero or negative order.



\end{document}